\documentclass[%
 aip,apl,
groupedaddress,
 amsmath,amssymb,
 reprint,%
]{revtex4-1}

\usepackage{graphicx}
\usepackage{dcolumn}
\usepackage{bm}

\usepackage[utf8]{inputenc}
\usepackage[T1]{fontenc}
\usepackage{amsthm}
\usepackage{color}
\usepackage{upgreek}
\newtheoremstyle{query}%
{}{}
{\color{red}}
{}
{\sffamily\bfseries}{:}{12pt}
{}
\theoremstyle{query}

\begin{document}

\preprint{AIP/123-QED}

\title[]{Electrical current switching of the noncollinear antiferromagnet Mn$_3$GaN}

\author{T. Hajiri}
 \email{t.hajiri@nagoya-u.jp}
\affiliation{Department of Materials Physics, Nagoya University, Nagoya 464-8603, Japan}
\author{S. Ishino}
\affiliation{Department of Materials Physics, Nagoya University, Nagoya 464-8603, Japan}
\author{K. Matsuura}
\affiliation{Department of Materials Physics, Nagoya University, Nagoya 464-8603, Japan}
\author{H. Asano}
\affiliation{Department of Materials Physics, Nagoya University, Nagoya 464-8603, Japan}
%

\date{\today}

\begin{abstract}
We report electrical current switching of noncollinear antiferromagnetic (AFM) Mn$_3$GaN/Pt bilayers at room temperature.
The Hall resistance of these bilayers can be manipulated by applying a pulse current of $1.5\times10^6$~A/cm$^2$, whereas no significant change is observed up to $\sim10^8$~A/cm$^2$ in Mn$_3$GaN single films, indicating that the Pt layer plays an important role. 
In comparison with ferrimagnetic Mn$_3$GaN/Pt bilayers, a lower electrical current switching of noncollinear AFM Mn$_3$GaN is demonstrated, with a critical current density two orders of magnitude smaller.
Our results highlight that a combination of a noncollinear AFM antiperovskite nitride and a spin-torque technique is a good platform of AFM spintronics.
\end{abstract}

\maketitle
Antiferromagnets (AFMs) are promising materials for spintronics application owing to a number of properties, such as the absence of stray magnetic fields, terahertz spin dynamics, and  stability against  external perturbations, and, in this context, electrical manipulation of the N\'eel vector is an important emerging technique.\cite{AFM_spintronics1, AFM_spintronics2, AFM_spintronics3, Romain_Nature}
Recent experiments have demonstrated electrical current switching of the AFM N\'eel vector by N\'eel spin--orbit torque (SOT) in CuMnAs\cite{CuMnAs_SOT, CuMnAs_SOT2} and Mn$_2$As,\cite{Mn2Au_SOT, Mn2Au_SOT2} and by spin torque in NiO/Pt bilayers\cite{PRL_NiO_SOT, Lorenzo_NiO_SOT} and Pt/NiO/Pt trilayers.\cite{Moriyama_NiO_SOT}
These experiments show that the AFM N\'eel vector can be manipulated by  electrical current densities of the order of $10^7$--$10^8$A/cm$^2$, which are similar to the current densities  required with SOT in typical ferromagnet (FM)/heavy metal (HM) bilayers.
On the other hand,  according to an early study of AFM spintronics,\cite{AFM_STT_theory1} a current density two orders of magnitude smaller may theoretically be capable of manipulating the AFM N\'eel vector.
Indeed, current-induced switching of exchange bias at such lower  current densities has been realized in AFM/FM bilayers.\cite{STT_exchange_bias1, STT_exchange_bias2, STT_exchange_bias3}
These results suggest the possibility of switching of the AFM N\'eel vector using lower electrical currents.

Recently, noncollinear AFMs become one of the most fascinating classes among AFMs, because they exhibit a large magneto-transport and a novel spin transport phenomena.\cite{Hua_Chen_PRL, orbital_AHE1, orbital_AHE2, noncollinear_spin}
However, the spin-torque switching of the AFM N\'eel vector is demonstrated in only collinear AFM, while the theoretical study shows that the domain of noncollinear AFM can be manipulated by spin current with realistic current density of $\sim10^7$A/cm$^2$.\cite{Yamane_noncollinear}
Thus, noncollinear AFMs call for the spin current injection thorough spin-Hall effect.
The antiperovskite manganese nitrides Mn$_3$GaN (MGN) is a good platform to investigate the electrical current manipulation of noncollinear AFMs, as well as for comparisons of the critical current density between FM and AFM materials, because MGN is a noncollinear AFM while a perpendicularly magnetized ferrimagnet can be obtained by applying tetragonal distortion.\cite{NIMS_MGN1, NIMS_MGN2, MGN_growth}
Both the N\'eel and Curie temperatures are reported to be above 380~K\cite{Neel_MGN1} and 660--740~K,\cite{NIMS_MGN1, NIMS_MGN2} respectively, which are relatively high  for this class of materials.

In the antiperovskite manganese nitrides Mn$_3A$N (where $A=$ Ni, Ga, Sn, etc.), the Mn atoms form a kagome lattice in the (111) plane.
These kagome lattices with nonzero Berry curvature may provide an interesting magneto-transport phenomena such as a large anomalous Hall effect (AHE) even with quite small canted magnetization of the order of $10^{-3}\mu_\mathrm{B}$ per atom, \cite{Hua_Chen_PRL, orbital_AHE1, orbital_AHE2} although the AHE is empirically proportional to net magnetization in conventional FMs.\cite{AHE1}
The AHE has been realized in the noncollinear AFMs Mn$_3X$ ($X$ = Sn, Ge, Pt) where the Mn atoms form a kagome lattice along different crystalline planes.\cite{Mn3Sn, Mn3Ge, Mn3Pt}
Since noncollinear AFM Mn$_3A$N has nonvanishing Berry curvature, the AHE can be realized.\cite{MGN_AHE, Cu-MNN_AHE, Mn3NiN_AHE}
This is an other motivation to study antiperovskite manganese nitrides.

In this letter, we investigate the spin-torque switching of noncollinear AFM MGN.
We observe significant changes in the Hall resistance when current pulses are applied to AFM-MGN/Pt bilayers at room temperature, while no clear change is observed in the longitudinal resistance.
By comparing AFM-MGN single films and FM-MGN/Pt bilayers, we show that the spin torque plays an important role in AFM-MGN/Pt bilayers, the critical current density for which is two orders of magnitude smaller than that for FM-MGN/Pt bilayers.
The possible origin of changes of the Hall resistance is discussed.

High-quality epitaxial AFM-MGN and FM-MGN films were prepared on MgO (001) substrates by reactive magnetron sputtering using a Mn$_{70}$Ga$_{30}$ target under an Ar/N$_2$ atmosphere. 
As reported in previous work,\cite{MGN_growth} AFM- and FM-MGN can be controllable by both the deposition rate and N$_2$ gass partial pressure.
The AFM-MGN films were grown at 450~$^{\circ}$C substrate temperature using 9~\%~N$_2$ + 91~\%~Ar gas mixtures with the deposition rate of 1.2~nm/min.
On the other hand, the FM-MGN films were grown at 450~$^{\circ}$C substrate temperature using 0.25~\%~N$_2$ + 99.75~\%~Ar gas mixtures with the deposition rate of 3.1~nm/min.
After the film growth was complete, we deposited a Pt layer (3~nm) by magnetron sputtering at room temperature.
The crystal structure was analyzed using out-of-plane X-ray diffraction (XRD) measurements with Cu~$K\alpha$ radiation.
The magnetic properties of the MGN/Pt bilayers were characterized using SQUID magnetometry. 
Transport properties were measured using Hall bars ($20~\upmu\mathrm{m}\times 100~\upmu\mathrm{m}$) prepared by a conventional photolithographic process.
The current is injected to MGN(bottom)/Pt(top) bilayers through the contact pads.

\begin{figure}[t]
\begin{center}
\includegraphics[width=\linewidth]{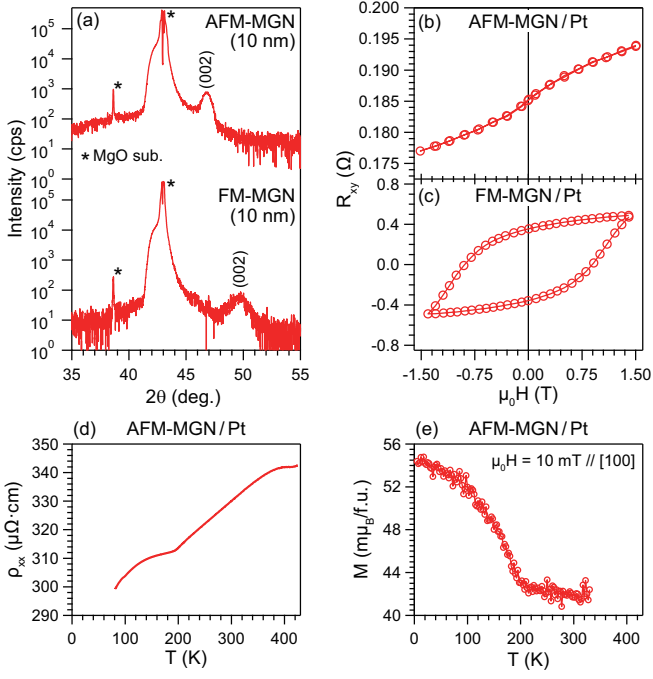}
\caption{
(a) Out-of-plane XRD profiles of 10~nm thick AFM-MGN and FM-MGN thin films. 
(b) and (c) AHE loops of (b) AFM-MGN/Pt and (c) FM-MGN/Pt bilayers.
(d) Temperature dependence of the resistivity of  AFM-MGN/Pt bilayers.
(e) Temperature dependence of the magnetization of AFM-MGN/Pt bilayers with an external magnetic field of 10~mT parallel to the [100] direction.}
\label{fig:one}
\end{center}
\end{figure}

Figure~\ref{fig:one}(a) shows the out-of-plane XRD patterns of  10~nm thick AFM-MGN and FM-MGN films on MgO substrates.
Only the (002) AFM- and FM-MGN peaks exhibit  Bragg peaks.
The out-of-plane lattice constants of AFM- and FM-MGN are 0.3885~nm and 0.3680
~nm, respectively.
The obtained out-of-plane lattice constants are consistent with those found for  AFM- and FM-MGN in previous studies.\cite{NIMS_MGN1, NIMS_MGN2, MGN_growth, MGN_growth, Neel_MGN1}
The AHE loops of AFM- and FM-MGN/Pt bilayers are shown in Figs.~\ref{fig:one}(b) and \ref{fig:one}(c), respectively.
A hysteresis loop is clearly observed for the FM-MGN/Pt bilayers.
For AFM-MGN/Pt bilayers, a nonlinear AHE curve is obtained, while no hysteresis loop is visible.
To determine the magnetic properties of AFM-MGN/Pt bilayers, the temperature dependences of the resistivity and magnetization were measured.
In the resistivity curve, there is a clear anomaly at 200~K and a kink at 380~K, as shown in Fig.~\ref{fig:one}(d).
It is well known that the antiperovskite nitrides Mn$_3A$N exhibit a kink at the AFM transition.\cite{Tashiro_MGN, Cu-MNN_AHE, antipero_RT}
In the bulk materials, several values of the N\'eel temperature have been reported, ranging from 280~K to more than 380~K.\cite{Neel_MGN1, Neel_MGN2}
Thus, the N\'eel temperature of our films can be determined to be  around 380~K.
On the other hand, in the temperature dependence of the magnetization shown in Fig.~\ref{fig:one}(e), an FM-like transition is observed at 200~K.
Since AFM-MGN/CoFe bilayers exhibit an exchange bias at low temperatures (not shown), we conclude that AFM and weak FM coexist below 200~K.

\begin{figure}[t]
\begin{center}
\includegraphics[width=\linewidth]{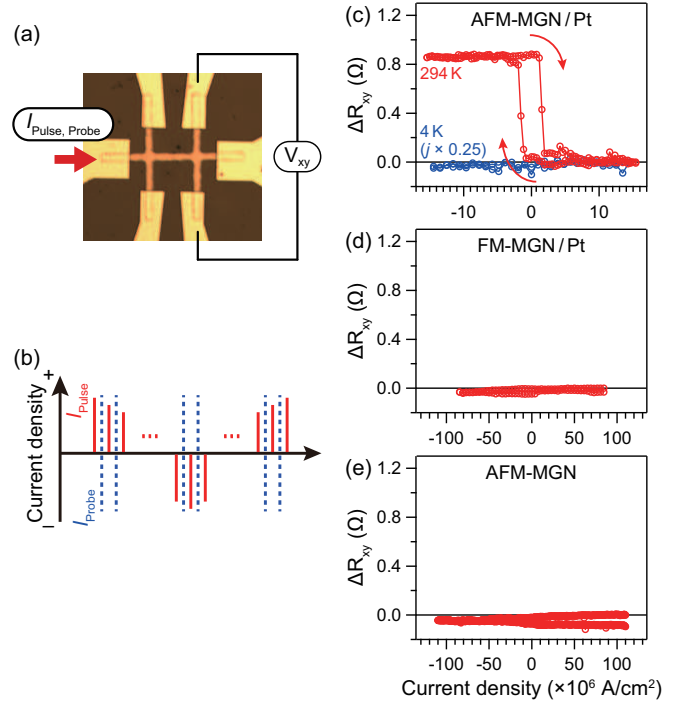}
\caption{
(a) Microscope image of AFM-MGN/Pt bilayers and experimental configuration.
(b) Electrical write--read sequence.
Between $I_\mathrm{Pulse}$ and $I_\mathrm{Probe}$, we set a delay of 4~s.
(c)--(e) Hall resistance $R_{xy}$ vs current density for (c) AFM-MGN/Pt bilayers, (d) FM-MGN/Pt bilayers, and (e) AFM-MGN films, respectively.
The current density $j$ at 4~K of AFM-MGN/Pt bilayers is multiplied by 0.25.
All measurements except those at 4~K for AFM-MGN/Pt bilayers were performed  at room temperature (294~K) without an external magnetic field.
}
\label{fig:two}
\end{center}
\end{figure}

To test the possibility of electrical current switching of noncollinear AFMs,  sequential write--read operations were performed.
A sequence of write--read operations is shown in Fig.~\ref{fig:two}(b).
After injection of a write current $I_\mathrm{Pulse}$ with 20~ms pulse width, the Hall resistance $R_{xy}$ was measured by a  probe current $I_\mathrm{Probe}$ of approximately $(0.5$--$1)\times 10^{5}$~A/cm$^2$.
Between $I_\mathrm{Pulse}$ and $I_\mathrm{Probe}$, we set a delay of 4~s.
All measurements presented here were performed without an external magnetic field.
The change in Hall resistance $\Delta R_{xy}$ as a function of  current density for AFM-MGN(10~nm)/Pt(3~nm) bilayers is shown in Fig.~\ref{fig:two}(c).
$R_{xy}$  changes sharply and a clear hysteresis loop is observed at  room temperature (294~K), indicating  success of the electrical write--read operation.
On the other hand, no change of $R_{xy}$ is observed at 4~K, suggesting that the appearance of weak FM may obstruct the electrical write operation in applied current density range.
The other possibility is that AHE signal discussed below disappears by the appearance of weak FM, so that no change of $R_{xy}$ is observed.
Figure~\ref{fig:two}(d) presents the current density dependence of $\Delta R_{xy}$ in FM-MGN(10~nm)/Pt(3~nm) bilayers in the absence of an external magnetic field.
It is well known that, in principle, an external magnetic field  parallel to the current is required for SOT in the case of a  perpendicularly magnetized FM, so no switching behavior is observed.
Thus, we also performed with an external magnetic field range of $-0.5\sim0.5$~T.
However, no significant switching was observed up to $\sim9.5\times10^7$~A/cm$^2$ in FM-MGN(10~nm)/Pt(3~nm) bilayers, indicating that a larger electrical current is needed.
Indeed, the critical current density is reported to be $(1.1$--$1.5)\times 10^{8}$~A/cm$^2$ in ferrimagnetic MnGa/HM bilayers.\cite{MnGa_SOT1, MnGa_SOT2}
To clarify the role of the Pt layer of AFM-MGN/Pt bilayers, we also performed the same operations on AFM-MGN(10~nm) single films.
As can be seen in Fig.~\ref{fig:two}(e), no current density dependence is observed in these single films.
This  indicates that the Pt layer plays an important role as a pure spin current generator through a spin Hall effect like the SOT in FM/HM bilayers.
From these results, we conclude that  spin-torque switching occurs in AFM-MGN/Pt bilayers.
From Fig.~\ref{fig:two}(c), the critical current density is estimated to be approximately $1.5\times 10^{6}$~A/cm$^2$, which is one to two orders of magnitude smaller than that for typical FM/HM bilayers\cite{CoFeB_SOT, Pt/Co_SOT} and the threshold current for NiO/Pt spin-torque switching.\cite{Lorenzo_NiO_SOT, Moriyama_NiO_SOT, PRL_NiO_SOT}
In the present study, Pt thickness is fixed to be 3~nm.
It is known that the thickness of the heavy metal layer can affect to the SOT in FM case,\cite{HM_thick_depe} so that it is worth to investigate the Pt layer thickness dependence in the future study.

\begin{figure}[t]
\begin{center}
\includegraphics[width=\linewidth]{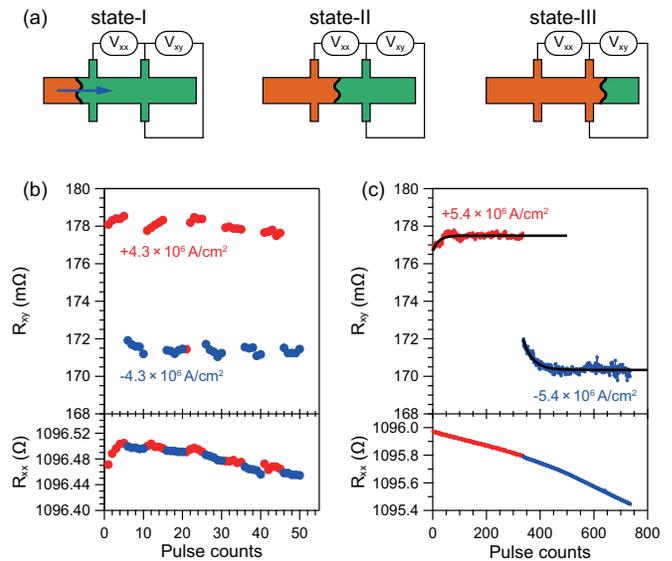}
\caption{
(a) Schematic illustration of AFM domain propagation.
(b) and (c) $R_{xy}$ and $R_{xx}$ vs pulse number at room temperature (294~K) of AFM-MGN/Pt using $I_\mathrm{Pulse}=\pm4.3\times 10^{6}$~A/cm$^2$ and $I_\mathrm{Pulse}=\pm5.4\times 10^{6}$~A/cm$^2$ of 20~ms width, respectively.
All measurements were performed at a probe current of  $I_\mathrm{Probe}$ of about $3\times 10^{5}$~A/cm$^2$ with 4~s delay from $I_\mathrm{Pulse}$ without an external magnetic field.
The bold solid lines in (c) are the  results of a fit to the exponential decay function $y=y_0+Ae^{-(x+x_0)/\tau}$. 
}
\label{fig:three}
\end{center}
\end{figure}

Next, we would like to discuss the possible cause of changes in $R_{xy}$.
In general, the Hall signals contain the ordinary Hall effect, the planar Hall effect, and the AHE.
The planar Hall effect is known to be one of  anisotropic magnetoresistance and is used in electrical detection of the AFM N\'eel order of CuMnAs,\cite{CuMnAs_SOT, CuMnAs_SOT2} Mn$_2$Au,\cite{Mn2Au_SOT, Mn2Au_SOT2} and IrMn.\cite{MnIr_AMR}
As  for FM SOT,\cite{SOT_domain} we assume that there are two types of AFM domains, which have high and low $R_{xy}$, respectively, and they can propagate under the influence of an electrical current, as shown in Fig.~\ref{fig:three}(a).
In fact, this  domain propagation is  expected theoretically even in noncollinear AFMs.\cite{Yamane_noncollinear}
If the planar Hall effect is dominant in our bilayers, then the $V_{xx, \mathrm{II/III}}$ signal will not be equal to $V_{xx, \mathrm{I}}$ and it will change when the $V_{xy}$ signals change.
On the other hand, if the AHE is dominant, then  the $V_{xx, \mathrm{II/III}}$ signal will be equal to $V_{xx, \mathrm{I}}$,\cite{Cu-MNN_AHE} while the $V_{xy, \mathrm{II/III}}$ signal will not be equal to $V_{xy, \mathrm{I}}$.
Therefore, we record $R_{xx}$ and $R_{xy}$ simultaneously to distinguish these two possibilities.

Figure~\ref{fig:three}(b) shows the pulse count dependence of $R_{xy}$ (top) and $R_{xx}$ (bottom) for AFM-MGN/Pt bilayers.
Here, $I_\mathrm{Pulse}=\pm4.3\times 10^{6}$~A/cm$^2$ with 20~ms pulse width was injected at each count.
After injection of $I_\mathrm{Pulse}$,  $R_{xy}$ was measured by a probe current $I_\mathrm{Probe}$ of about $3\times 10^{5}$~A/cm$^2$ with a delay of 4~s from $I_\mathrm{Pulse}$.
This procedure is the same as in Fig.~\ref{fig:two}, although a  different bilayer was measured, so $\Delta R_{xy}$ is different.
We note that  almost the same critical current density was obtained for the several different AFM-MGN/Pt bilayers we studied, but $\Delta R_{xy}$ shows a large sample dependence, probably because of local variations in the quality of the AFM-MGN film.
$R_{xy}$ jumps between $178.05\pm0.06$~m$\Omega$ and $171.34\pm0.04$~m$\Omega$ on $\pm I_\mathrm{Pulse}$ injection, except for a single error at the 21st injection.
The obtained $\Delta R_{xy}$ and $\Delta R_{xy}$ ratio are 6.71~m$\Omega$ and 3.84~$\%$, respectively, which are enough larger than the experimental errors.
At the same time, $R_{xx}$ shows small oscillations but no clear jumps on $\pm I_\mathrm{Pulse}$ injection.
If the domains propagate from state I to state II, then $R_{xx}$ will increase/decrease monotonically  with respect to the domain wall position.
On the other hand, if the domains propagate from state I to state III, then $R_{xx}$ will  jump between two values like $R_{xy}$.
Since no such increasing/decreasing or jump behavior of $R_{xx}$ is observed on $\pm I_\mathrm{Pulse}$ injection, the planar Hall effect can be excluded as a cause of $\Delta R_{xy}$.

The AHE in noncollinear AFMs is theoretically expected and experimentally realized in antiperovskite Mn$_3$NiN systems.\cite{Cu-MNN_AHE, Mn3NiN_AHE}
Although AFM-MGN is also theoretically expected to exhibit the AHE,\cite{MGN_AHE} there have been no reports of this to date.
As shown in Fig.~\ref{fig:one}(b), a nonlinear AHE curve is obtained for AFM-MGN/Pt bilayers, showing that the AHE possibly occurs for these.
These results may suggest that the magnetic anisotropy of AFM-MGN is too large to manipulate by the application of an external magnetic field, which deserves future study.

After several hundred cycles of $\pm I_\mathrm{Pulse}$ injection, we finally performed  continuous $+I_\mathrm{Pulse}$ or $-I_\mathrm{Pulse}$ write operations.
The results for AFM-MGN/Pt bilayers are shown in Fig.~\ref{fig:three}(c).
The $R_{xy}$ switching still survives even after several hundred cycles of write--read operations.
$R_{xy}$ increases (decreases) on continuous $+I_\mathrm{Pulse}$ ($-I_\mathrm{Pulse}$) injection and then exhibits saturation.
As discussed for SOT of synthetic AFMs,\cite{synthetic_AFM_SOT} the observed asymptotic $R_{xy}$ behavior can be fitted by an exponential decay function $y=y_0+Ae^{-(x+x_0)/\tau}$.
The time constants $\tau$ for $\pm I_\mathrm{Pulse}$ injection are different: $\tau=21.7$ for $+I_\mathrm{Pulse}$ and 35.6 for $-I_\mathrm{Pulse}$.
This behavior has not been reported for other AFM/HM bilayers.\cite{Lorenzo_NiO_SOT, synthetic_AFM_SOT}
$R_{xx}$ was also recorded at the same time, as shown at the bottom of Fig.~\ref{fig:three}(c).
However, since the room temperature  decreased monotonically during the continuous $\pm I_\mathrm{Pulse}$ injection measurements, so did $R_{xx}$.
On the other hand, there is no clear relationship between $R_{xy}$ and decreasing room temperature, showing the stability against  temperature perturbations.
Together with the stability against write operations, these results highlight the potential application of this approach.

In summary, we have shown the possibility of spin-torque switching of noncollinear AFM-MGN at room temperature.
A significant change in Hall resistance is observed on the application of current pulses to AFM-MGN/Pt bilayers, while no clear change is observed in AFM-MGN single films, indicating that spin torque plays an important role in manipulating the Hall resistance.
We have demonstrated that the critical current density for AFM-MGN is two orders of magnitude smaller than that for FM-MGN and that  $R_{xy}$ switching can survive even after several hundred cycles of write--read operations.
By simultaneous recording of $R_{xx}$ and $R_{xy}$ as  functions of applied pulse counts, possible evidence of the AHE in AFM-MGN has been found.
These results open the pathway to the efficient control of noncollinear AFM order.


\begin{acknowledgments}
This work was supported by the Japan Society for the Promotion of Science (JSPS) KAKENHI Grant No. 17K17801 and 17K19054, a research grant from the Murata Science Foundation, Kyosho Hatta Foundation, and the Center for Spintronics Research Network (CSRN) of Tohoku University.
T.~H. also acknowledges a Research Grant for Young Scientists (Research Center for Materials Backcasting Technology, School of Engineering, Nagoya University).
\end{acknowledgments}


\section*{REFERENCES}
\begin{thebibliography}{99}
\bibitem{AFM_spintronics1} H. V. Gomonay and V. M. Loktev, Low Temp. {\bf 40}, 17 (2014).
\bibitem{AFM_spintronics2} T. Jungwirth, X. Mart\'{i}, P. Wadleyand, and J. Wunderlich, Nat. Nanotechnol. {\bf 11}, 231 (2016).
\bibitem{AFM_spintronics3} V. Baltz, A. Manchon, M. Tsoi, T. Moriyama, T. Ono, and Y. Tserkovnyak, Rev. Mod. Phys. {\bf 90}, 015005 (2018).
\bibitem{Romain_Nature} R. Lebrun, A. Ross, S. A. Bender, A. Qaiumzadeh, L. Baldrati, J. Cramer, A. Brataas, R. A. Duine and M. Kl\"{a}ui, Nature {\bf 561}, 222 (2018).
\bibitem{CuMnAs_SOT} P. Wadley, B. Howells, J. \v{Z}elezn\'{y}, C. Andrews, V. Hills, R. P. Campion, V. Nov\'{a}k, K. Olejnik, F. Maccherozzi, S. S. Dhesi, S. Y. Martin, T. Wagner, J. Wunderlich, F. Freimuth, Y. Mokrousov, J. Kune\v{s}, J. S. Chauhan, M. J. Grzybowski, A. W. Rushforth, K. W. Edmonds, B. L. Gallagher, and T. Jungwirth, Science {\bf 351}, 587 (2016).
\bibitem{CuMnAs_SOT2} T. Matalla-Wagner, M.-F. Rath, D. Graulich, J.-M. Schmalhorst, G. Reiss, and M. Meinert, eprint arXiv:1903.12387 (2019).
\bibitem{Mn2Au_SOT} S. Y. Bodnar, L. \v{S}mejkal, I. Turek, T. Jungwirth, O. Gomonay, J. Sinova, A. A. Sapozhnik, H. J. Elmers, M. Kl\"{a}ui, and M. Jourdan, Nat. Commun. {\bf 9}, 348 (2018). 
\bibitem{Mn2Au_SOT2} M. Meinert, D. Graulich, and T. Matalla-Wagner, Phys. Rev. Appl. {\bf 9}, 064040 (2018).
\bibitem{PRL_NiO_SOT} X. Z. Chen, R. Zarzuela, J. Zhang, C. Song, X. F. Zhou, G. Y. Shi, F. Li, H. A. Zhou, W. J. Jiang, F. Pan, and Y. Tserkovnyak, Phys. Rev. Lett. {\bf 120}, 207204 (2018).
\bibitem{Lorenzo_NiO_SOT} L. Baldrati, O. Gomonay, A. Ross, M. Filianina, R. Lebrun, R. Ramos, C. Leveille, T. Forrest, F. Maccherozzi, E. Saitoh, J. Sinova, and M. Kl\"{a}ui, eprint arXiv:1810.11326 (2018).
\bibitem{Moriyama_NiO_SOT} T. Moriyama, K. Oda, and T. Ono, Sci. Rep. {\bf 8}, 14167 (2018).
\bibitem{AFM_STT_theory1} A. S. N\'{u}\~{n}ez, R. A. Duine, Paul Haney, and A. H. MacDonald, Phys. Rev. B {\bf 73}, 214426 (2006).
\bibitem{STT_exchange_bias1} Z. Wei, A. Sharma, A. S. Nunez, P. M. Haney, R. A. Duine, J. Bass, A. H. MacDonald, and M. Tsoi, Phys. Rev. Lett. 98, 116603 (2007).
\bibitem{STT_exchange_bias2} X.-L. Tang, H.-W. Zhang, H. Su, Z.-Y. Zhong, and Y.-L. Jing, Appl. Phys. Lett. {\bf 91}, 122504 (2007).
\bibitem{STT_exchange_bias3} H. Sakakibara, H. Ando, Y. Kuroki, S. Kawai, K. Ueda, and H. Asano, J. Appl. Phys. {\bf 117}, 17D725 (2015).

\bibitem{Hua_Chen_PRL} H. Chen, Q. Niu, and A. H. MacDonald, Phys. Rev. Lett. {\bf 112}, 017205 (2014).
\bibitem{orbital_AHE1} T. Tomizawa and H. Kontani, Phys. Rev. B {\bf 80}, 100401 (2009). 
\bibitem{orbital_AHE2} T. Tomizawa and H. Kontani, Phys. Rev. B {\bf 82}, 104412 (2010).
\bibitem{noncollinear_spin} J.  \v{Z}elezn\'{y}, Y. Zhang, C. Felser, and B. Yan, Phys. Rev. Lett. {\bf 119}, 187204 (2017).
\bibitem{Yamane_noncollinear} Y. Yamane, O. Gomonay, and J. Sinova, eprint arXiv:1901.05684 (2019).

\bibitem{NIMS_MGN1} H. Lee, H. Sukegawa, J. Liu, T. Ohkubo, S. Kasai, S. Mitani, and K. Hono, Appl. Phys. Lett. {\bf 107}, 032403 (2015).
\bibitem{NIMS_MGN2} H. Lee, H. Sukegawa, J. Liu, S. Mitani, and K. Hono, Appl. Phys. Lett. {\bf 109}, 152402 (2016).
\bibitem{MGN_growth} S. Ishino, J. M. So, H. Goto, T. Hajiri, and H. Asano, AIP Adv. {\bf 8}, 056312 (2018).
\bibitem{Neel_MGN1} W. J. Feng, D.Li, Y. F. Dend, Q. Zhang and H. H. Zhang, J. Mater. Sci. {\bf 45}, 2770 (2010).
\bibitem{AHE1} N. Nagaosa, J. Sinova, S. Onoda, A. H. MacDonald, and N. P. Ong, Rev. Mod. Phys. {\bf 82}, 1539 (2010).
\bibitem{Mn3Sn} S. Nakatsuji, N. Kiyohara, and T. Higo, Nature {\bf 527}, 212 (2015).
\bibitem{Mn3Ge} N. Kiyohara, T. Tomita, and S. Nakatsuji, Phys. Rev. Appl. {\bf 5}, 064009 (2016).
\bibitem{Mn3Pt} Z. Liu, H. Chen, J. Wang, J. Liu, K. Wang, Z. Feng, H. Yan, X. Wang, C. Jiang, J. Coey \emph{et al.}, Nat. Electron.{\bf 1}, 172 (2018).
\bibitem{Cu-MNN_AHE} K. Zhao, T. Hajiri, H. Chen, R. Miki, H. Asano, and P. Gegenwart, eprint arXiv:1904.05678 (2019).
\bibitem{Mn3NiN_AHE} D. Boldrin, I. Samathrakis, J. Zemen, A. Mihai, B. Zou, B. Esser, D. McComb, P. Petrov, H. Zhang, and L. F. Cohen, eprint arXiv:1902.04357 (2019).
\bibitem{MGN_AHE} G. Gurung, D.-F. Shao, T. R. Paudel, and E. Y. Tsymbal, Phys. Rev. Mater. {\bf 3}, 044409 (2019).

\bibitem{Tashiro_MGN} H. Tashiro, R. Suzuki, T. Miyawaki, K. Ueda, and H. Asano, J. Korean Phys. {\bf 63}, 299 (2013).
\bibitem{antipero_RT} M. Hadano, A. Ozawa, K. Takenaka, N. Kaneko, T. Oe, and C. Urano, J. Appl. Phys. {\bf 111}, 07E120 (2012).
\bibitem{Neel_MGN2} D. Kasugai, A. Ozawa, T. Inagaki, and K. Takenaka, J. Appl. Phys. {\bf 111}, 07E314 (2012).
\bibitem{MnGa_SOT1} K. K. Meng, J. Miao, X. G. Xu, Y. Wu, X. P. Zhao, J. H. Zhao, and Y. Jiang, Phys. Rev. B {\bf 94}, 214413 (2016).
\bibitem{MnGa_SOT2} K. K. Meng, J. Miao, X. G. Xu, Y. Wu, J. X. Xiao, J. H. Zhao and Y. Jiang, Sci. Rep. {\bf 6}, 38375 (2016).
\bibitem{CoFeB_SOT} L. Liu, C.-F. Pai, Y. Li, H. W. Tseng, D. C. Ralph, and R. A. Buhrman,  Science {\bf 336}, 555 (2012).
\bibitem{Pt/Co_SOT} L. Liu, O. J. Lee, T. J. Gudmundsen, D. C. Ralph, and R. A. Buhrman, Phys. Rev. Lett. {\bf 109}, 096602 (2012).

\bibitem{HM_thick_depe} S. Cho, S. C. Baek, K.-D. Lee, Y. Jo, and B. G. Park, Sci. Rep. {\bf 5}, 14668 (2015).

\bibitem{MnIr_AMR} X. Zhou, L. Ma, Z. Shi, W. J. Fan, R. F. L. Evans, J.-G. Zheng, R. W. Chantrell, S. Mangin, H. W. Zhang, and S. M. Zhou, Sci. Rep. {\bf 5}, 9183 (2015).
\bibitem{SOT_domain} P. Sethi, S. Krishnia, W. L. Gan, F. N. Kholid, F. N. Tan, R. Maddu and W. S. Lew, Sci. Rep. {\bf 7}, 4964 (2017).

\bibitem{synthetic_AFM_SOT} T. Moriyama, W. Zhou, T. Seki, K. Takanashi, and T. Ono, Phys. Rev. Lett. {\bf 121}, 167202 (2018).

\end{thebibliography}
\end{document}